# Atomic-resolution TEM Studies of Pillar-Matrix Structures in Epitaxially Grown Ultrathin $ZrO_2$-$La_{2/3}Sr_{1/3}MnO_3$ Films

Dan Zhou[1*], Wilfried Sigle[1], Eiji Okunishi[2], Yi Wang[1], Marion Kelsch[1], Hanns-Ulrich Habermeier[3], and Peter A. van Aken[1]

[1]Max Planck Institute for Intelligent Systems, Stuttgart Center for Electron Microscopy, Heisenbergstrasse 3, 70569, Stuttgart, Germany

[2]JEOL Ltd., 1-2 Musashino, 3-Chome Akishima, 196-8558, Tokyo, Japan

[3]Max Planck Institute for Solid State Research, Heisenbergstrasse 1, 70569, Stuttgart, Germany

[*]E-mail: danzhou@is.mpg.de

**ABSTRACT**: We studied $ZrO_2$-$La_{2/3}Sr_{1/3}MnO_3$ pillar–matrix thin films which were found to show anomalous magnetic and electron transport properties controlled by the amount of $ZrO_2$. With the application of an aberration–corrected transmission electron microscope, structure and chemical information of the system, especially of the pillar–matrix interface were revealed at atomic resolution. Minor amounts of Zr were found to occupy Mn positions within the matrix and its solubility within the matrix was found to be less than 6 mol%. Moreover, the Zr concentration reached minimum concentration at the pillar–matrix interface accompanied by oxygen deficiency. La and Mn diffusion into the pillar was observed along with a change of the Mn valence state. La and Mn positions inside $ZrO_2$ pillars were also revealed at atomic resolution. These results provide detailed information for future studies of macroscopic properties of these materials.

**I INTRODUCTION**

In functional oxides such as superconducting cuprates and ferromagnetic manganites the investigation of electrically insulating columnar oxide defects embedded in the conducting matrix materials has experienced an increasing research activity in the past decade.[1-4] From the fundamental point of view, this activity was initiated by the exploration of the crystalline structure and elemental distribution at the interface between the columnar defects and the matrix material with the subsequent strain- and defect-induced modification of their electronic and magnetic properties. Elemental and electronic reconstructions at the interface between complex oxides with different functionalities are challenging phenomena to be explored. They give rise to interference effects between the states across interfaces with the potential to generate new properties and functionalities.[5, 6] Research on self-assembled vertically aligned nanocomposite thin films with two immiscible components hetero-epitaxially grown on single crystal substrates represents another branch of these activities.[1-3, 7] These structures have the advantage of utilizing the functionalities of both components with the possibility to tune the material properties by tailoring the interface-to-volume ratio, hetero-epitaxial strain, or modifying the cation valence state. From the application point of view, columnar non-conducting $BaZrO_3$ or $SrZrO_3$ defects in superconducting $YBa_2Cu_3O_7$ thin films have been proven to enhance the flux-line pinning properties drastically and are used in several 3$^{rd}$ generation commercial superconducting tape fabrication technologies.[8, 9] Additionally, the thermoelectric figure of merit $ZT = (S^2\sigma/\kappa)T$ (with $S$ being the Seebeck-coefficient, $\sigma$ and $\kappa$ the electrical and thermal conductivity, respectively) can be enlarged by introducing precipitates and columnar defects, thus reducing the phonon part of the thermal conductivity in order to achieve an "electron crystal / phonon glass''–type material.[10, 11]

In ultrathin $La_{2/3}Sr_{1/3}MnO_3$ (LSMO) films ( thickness ~ 45 nm ) with a low density ( 0 – 6 mol% ) of $ZrO_2$ precipitates an unusual low temperature resistivity increase associated with quantum interference effects of the electron waves was observed in conjunction with anomalous magnetic anisotropies.[12-14] Evidently, electronic transport of LSMO is sensitively influenced by the presence of $ZrO_2$ precipitates. Here, we report on the structure and composition of these precipitates, which are found to exist as pillars in the LSMO matrix. We use aberration-corrected scanning transmission electron microscopy (STEM) and simultaneous electron energy-loss spectroscopy (EELS) to reveal the structure, composition, and valence state at atomic resolution[15-18] especially for the pillar–matrix interface.

## II EXPERIMENTAL SECTION

Zirconium oxide ($ZrO_2$) and lanthanum strontium manganese oxide ($La_{2/3}Sr_{1/3}MnO_3$, LSMO) were co-deposited epitaxially on (001) single-crystalline lanthanum aluminum oxide ($LaAlO_3$, LAO) substrate by pulsed laser deposition. Stoichiometric amounts of LSMO and $ZrO_2$ according to (1-$x$)LSMO+ $x$$ZrO_2$, with $x$= 0, 0.06, 0.2, and 0.3, were used. Details of the material growth process can be found in.[12]

In order to obtain 3-dimensional information, specimens for transmission electron microscopy (TEM) studies were prepared perpendicular (cross-sectional view) and parallel (plan-view) to the substrate by grinding, dimpling, and Argon ion thinning with a precision ion polishing system (PIPS, Gatan, model 691) under liquid nitrogen cooling to achieve electron transparency.

High-angle annular dark-field (HAADF) images, electron energy-loss spectroscopy (EELS) spectrum images from plan-view specimens were obtained using a probe-aberration-corrected JEOL ARM200CF microscope operated at 200 keV, equipped with a Gatan GIF Quantum ERS imaging filter with dual-EELS acquisition capability. The experimental convergence angle was 30 mrad for HAADF and EELS imaging. The corresponding inner and outer collection semi-angles for HAADF were set to 54-220 mrad. The inner and outer collection semi-angles for annular dark field (ADF) images acquired simultaneously during EELS spectrum imaging with Gatan ADF detector were 72−172 mrad, and the collection angle for EELS spectrum imaging was 72 mrad. Multivariate statistical analysis (MSA) was performed to reduce the noise of the EEL spectra with weighted principle-component analysis (PCA). From HAADF images dislocations have been identified and strain distributions were calculated using geometric phase analysis (GPA) software from HREM Research Inc..

## III RESULTS AND DISCUSSION

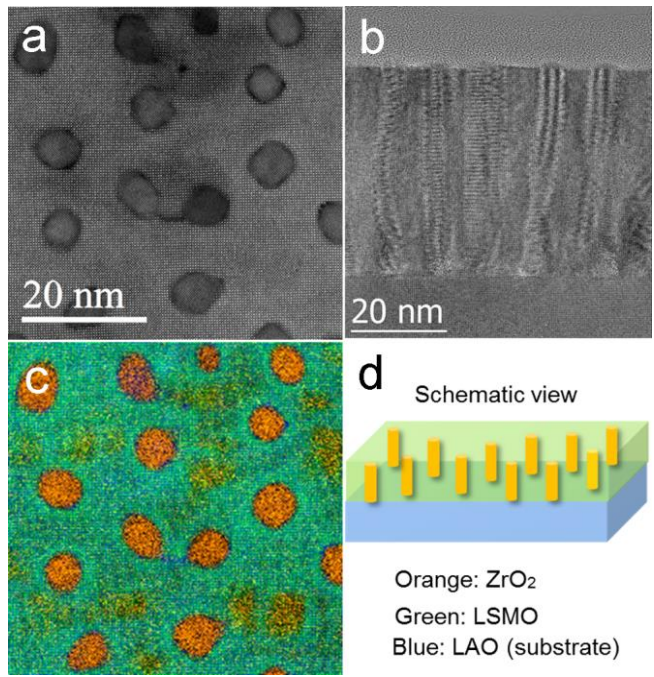

FIG. 1. (a) Annular dark field (ADF) image of a plan-view 70 mol%LSMO-30 mol%ZrO$_2$ sample; (b) HREM image of the side-view specimen with 30 mol% ZrO$_2$; (c) Three-color EELS spectrum image of the area shown in (a) with Zr-L$_{2,3}$ in orange, Mn-L$_{2,3}$ in blue and La-M$_{4,5}$ in green; (d) A schematic view of the sample.

The pillar structure is verified for both plan-view and side-view specimens. Fig. 1(a) shows the top-view HAADF image of the ZrO$_2$(30 mol%)−LSMO thin film with clear circular precipitates. The aberration-corrected HREM image [Fig. 1(b)] of the side-view specimen reveals a columnar structure within the film and the film thickness is measured to be about 45 nm. The phase separation manifests itself in a Moiré contrast arising from the overlap of ZrO$_2$ and LSMO crystal lattices having different lattice parameters. It shows that the pillars extend all the way through the LSMO film until the substrate. Even though all the pillars penetrate through the film thickness, some penetrate straight with a relatively sharp edge, while other pillars are bent.

The EELS spectrum image of Fig. 1(a) is shown in Fig. 1(c) where the intensities of Zr-L$_{2,3}$, La M$_{4,5}$, and Mn-L$_{2,3}$ edges are displayed in orange, green, and blue, respectively. Oxygen is distributed everywhere and is not shown here. It is clear that pillar-shaped structures of ZrO$_2$ have formed. They show circular or elliptical, sometimes faceted circumferences. A schematic view of the whole specimen is presented in Fig. 1(d).

From a statistical analysis of the pillar areas a diameter of (5.1 ± 0.8) nm was deduced for the 30 mol% ZrO$_2$ specimen and of (3.7 ± 0.8) nm for the 20 mol% ZrO$_2$ sample.

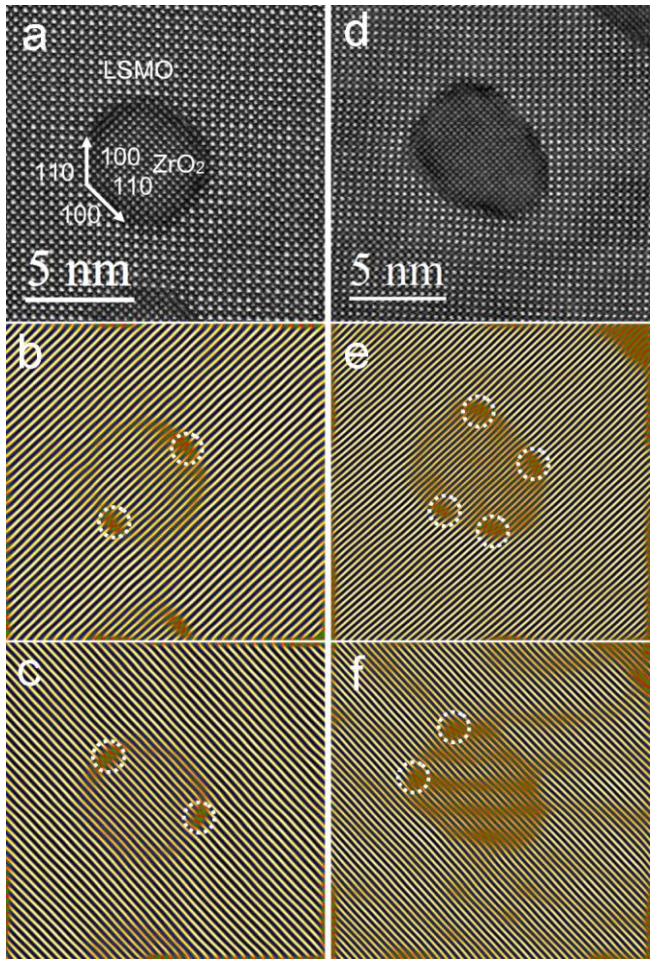

FIG. 2. Fourier filtered image using (b) {100} reflection and (c) {010} reflection of the HAADF image of a plan-view 70 mol%LSMO-30 mol%ZrO$_2$ sample shown in (a): And fourier filtered image using (e) {100} reflection and (f) {010} reflection of the HAADF image of a plan-view 70 mol%LSMO-30 mol%ZrO$_2$ sample shown in (d).

Fig. 2(a) is a high magnification HAADF image of a pillar with sharp interfaces. In the matrix region, brighter columns correspond to heavy La/Sr (A-site) ions, and weaker columns correspond to the lighter mixed Mn-O (B-site) ions. In the pillar region, only Zr columns are visible because oxygen columns are invisible in HAADF due to the small scattering cross section. Towards the interface to the matrix the column brightness is reduced. As will be explained later, this is most likely related to the substitution of Zr by Mn.

As a [001]-oriented LAO substrate was used, the epitaxial relationship between the thin film and the substrate can be determined from the HAADF and HREM image of the top-view and cross-section specimens. To facilitate the correlation of crystallographic orientation between LAO, LSMO, and $ZrO_2$, we use here LAO[19] and LSMO[20, 21] in the tetragonal notation (Space group I4/mcm, No. 140) and $ZrO_2$ in the tetragonal system (space group P42/NMCS, No. 137).[22] The corresponding crystal structure can be found in Fig. S1. With our low sample-preparation temperature of 800 °C,[12] pure $ZrO_2$ is expected to crystallize in the monoclinic phase[23, 24] which however doesn't match with the square atomic arrangement seen in the HAADF image [Fig. 2(a)]. We believe that tetragonal or cubic $ZrO_2$ is formed by partial substitution of Zr by Mn atoms as will be explained in detail later. Here we describe it in the tetragonal system to facilitate our description. Thus the epitaxial relationships of LSMO and $ZrO_2$ with respect to the LAO substrate are as follows:

LSMO[100]//LAO[100], LSMO[001]//LAO[001], $ZrO_2$[110]// LSMO[100]

The presence of faceted interfaces between $ZrO_2$ and LSMO [Fig. 2(a)] indicates a preference towards $[110]_{ZrO2}/[100]_{LSMO}$, and $[100]_{ZrO2}/[110]_{LSMO}$ interfaces.

As reported in a previous paper on this group of material,[14] the a and b values of LSMO are 5.4472 Å, 5.4483 Å, 5.4624 Å, and 5.4471 Å for LSMO films with 0 mol%, 3 mol%, 6 mol%, and 20 mol% $ZrO_2$, respectively. This means that the LSMO crystal lattice expands by the addition of $ZrO_2$, but partially relaxes again at high $ZrO_2$ content.

We use the *a* and *b* values of LSMO with 20 mol% $ZrO_2$. For $ZrO_2$, depending on sample preparation, a values range from 3.562 Å[22] to 3.646 Å.[25] Thus LSMO and $ZrO_2$ is expected to have a mismatch of about (5.3–7.5)%. For the 30 mol% $ZrO_2$ sample, by measuring the plane distances of $ZrO_2$ pillars and surrounding matrix, we found a lattice mismatch between LSMO and $ZrO_2$ of from 5.6 % to 9.2%. These values are large enough to drive nucleation of misfit dislocations at the interfaces, as is confirmed by TEM. The visibility of misfit dislocations in the ab-plane can be enhanced by Fourier filtering selecting {100} and {010} reflections in the GPA analysis [Fig. 2]. For the pillar shown in Fig. 2(a), the misfit dislocations appear in pairs along a and b planes, as shown in Fig. 2(b), (c). Fig. 2(e), (f) show the dislocations along a and b planes for the pillars shown in Fig. 2(d). It is directly visible that the dislocations along b planes are not paired. Choosing the $x$-axis parallel to [100] and $y$-axis parallel to [010] direction and using the matrix as reference area, we get the symmetric strain-field image in Fig. S2. It is clear that the compressive strain stays in the matrix regions with non-paired misfit dislocations, and strain relaxed for the matrix regions with paired misfit dislocations.

These examples show that the overall structure of the pillars has not reached elastic equilibrium yet. This is because, according to the lattice misfit, the pillar size has reached a value which only just enables nucleation of a misfit dislocation. Probably because of insufficient thermal activation (low deposition temperature, 770 °C[12]) or high deposition rate, not every pillar has managed to nucleate a sufficient number of misfit dislocations and remains in a 'superstrained' state with tensile stresses on the $ZrO_2$ side of the interface. The formation of misfit dislocations is one way to relax the strain in a two-component system. Another way to relax the strain is to allow for interdiffusion. This reduces the abruptness of the interfaces in terms of lattice misfit and chemical potential.[26] In the present system we found several evidences for interdiffusion which will be discussed below: (i) Zr atoms are detected in the LSMO matrix; (ii) Mn atoms are found within the $ZrO_2$ pillars; (iii) a third Mn-rich phase is formed connecting adjacent pillars, which will however be presented in another separate paper.

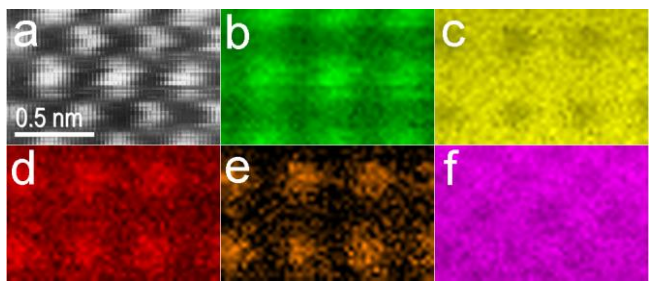

FIG 3. EELS spectrum image of (b) La-$M_{4,5}$, (c) Sr-$L_{2,3}$, (d) Mn-$L_{2,3}$, (e) Zr-$L_{2,3}$, (f) O-K of the LSMO matrix area in a plan-view 70 mol%LSMO-30 mol%$ZrO_2$ sample shown in the ADF image in (a).

The atomic resolution EELS spectrum imaging (SI) of the matrix region in the 30 mol%-$ZrO_2$ sample (processed with PCA) is shown in Fig. 3. La and Sr [Fig. 3(b) and Fig. 3(c)] occupy the same locations as expected from the LSMO structure. Mn and Zr [Fig 3(d) and Fig. 3(e)] take the same location which confirms that Zr is present in the matrix and occupies Mn/O column positions. The O map shows minimum values at La/Sr positions while being present everywhere else, which is as expected from the LSMO structure.

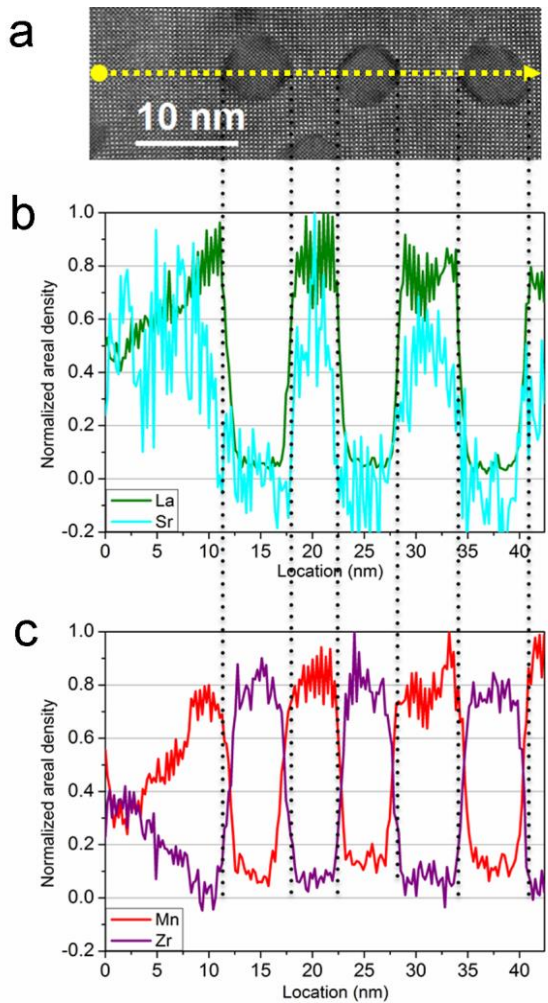

FIG. 4. EELS spectrum line profile of (b) La-$M_{4,5}$ and Sr-$L_{2,3}$, and (c) Mn-$L_{2,3}$ and Zr-$L_{2,3}$ of the line drawn in the HAADF image of a plan-view 70 mol%LSMO-30 mol%$ZrO_2$ sample in (a).

Elemental concentration profiles across the $ZrO_2$/LSMO interface are not abrupt. EELS SI in Fig. 4 shows a concentration line profile from EELS data across a few pillars, indicating that La and Mn concentrations increase when approaching the pillar from the matrix side, and quickly drop after entering the pillar region, however not reaching zero concentration. This shows that La and Mn atoms are present in the zirconia lattice. Both Sr and Zr concentrations decrease when approaching the pillar from the matrix side. Sr is absent inside the pillar, Zr concentration reaches a minimum close to zero at just in front of the interface, which will be explained later. These concentration gradients are much less pronouced or even absent in regions where pillars are close to each other, as in the right part of Fig. 4.

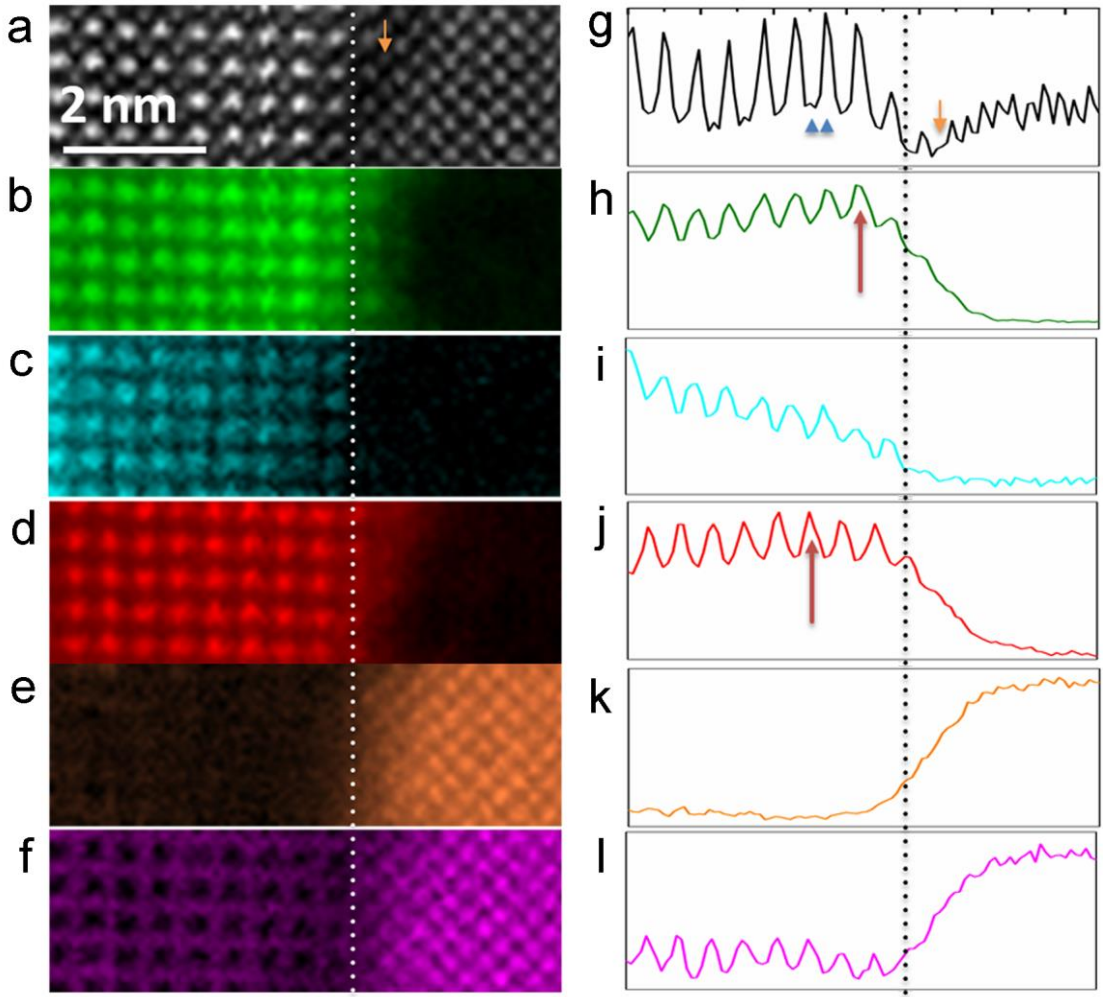

FIG. 5. Atomic resolution EELS spectrum image of (b) La-$M_{4,5}$, (c) Sr-$L_{2,3}$, (d) Mn-$L_{2,3}$, (e) Zr-$L_{2,3}$, and (f) O-K of the pillar-matrix interface area shown in (a) ADF image of a plan-view 70 mol%LSMO-30 mol%$ZrO_2$ sample: and integrated line profile of images (a) to (f) shown in (g) to (l) respectively.

Detailed elemental distribution can be revealed by atomically resolved EELS maps as displayed in Fig. 5 along with the simultaneously acquired ADF image. In the bulk regions, the elements are distributed as expected from the structure models of LSMO and $ZrO_2$: Within LSMO, La [Fig. 5(b)] and Sr columns [Fig. 5(c)] take the same positions, Mn-O [Fig. 5(d) & (f)] columns are located in the center of four La/Sr columns, and pure oxygen [Fig. 5(f)] is located between La/Sr positions. Mn is octahedrally surrounded by oxygen atoms. Along the <001>-projection, mixed Mn−O columns and pure O columns exist. In the Mn map the Mn—O columns are clearly resolved. Because of the limited spatial resolution of the oxygen map [Fig. 5(f)] the columns containing oxygen are not visible separately but appear as horizontal and vertical stripes in LSMO. Within $ZrO_2$, Zr [Fig. 5(e)] and O [Fig. 5(f)] columns are resolved. La and Mn are found to exist within the pillar, while the concentrations are too low to be atomically resolved by this analysis. They will be revealed by Gaussian fitting later.

From the atomically resolved EELS maps and the HAADF image, line profiles were obtained by integrating intensities in the direction parallel to the interface. These are displayed Fig. 5 (g-l). We define the interface position where the Sr signal vanishes (dotted line). For the HAADF image, we find an increasing intensity followed by a damping on the LSMO side when approaching the interface. Such a damping is also visible on the $ZrO_2$ side of the interface. This damping is obvious in every HAADF image as a dark circular region around each pillar. Since HAADF image intensity is proportional to the atomic number, the intensity damping here can be directly correlated with the elemental distribution at the interface.

Oxygen deficiency was found to be shown on both LSMO and $ZrO_2$ sides of the interface, as can be seen from Fig. 5 (f & l).

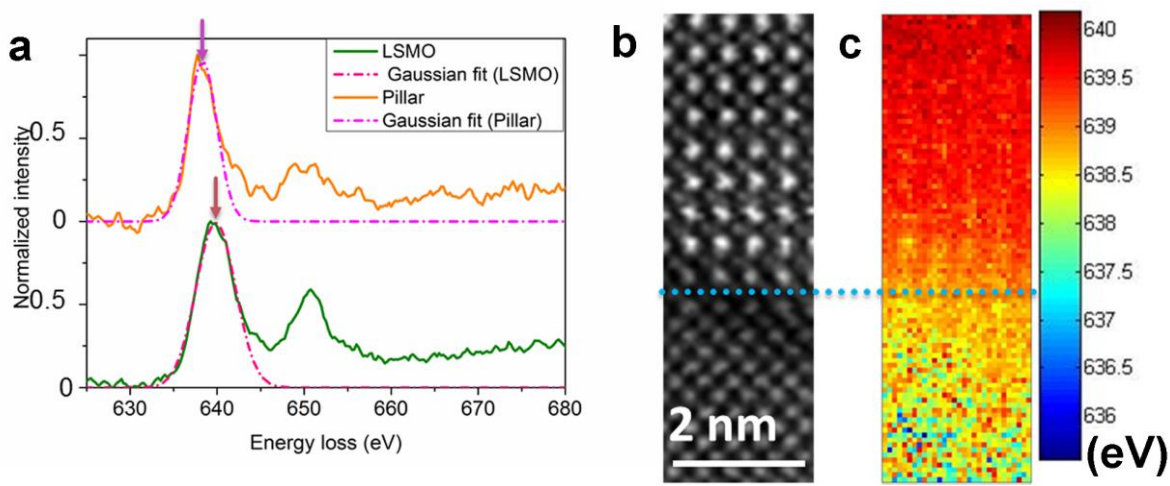

FIG. 6. (a) Nonlinear least square fitting with Gaussian function on Mn-$L_3$ edge peak position of the two spectra from pillar and matrix region of the plan-view 70 mol%LSMO-30 mol%$ZrO_2$ sample respectively; (b) Replica of Fig. 5(a), which is turned 90 degree clockwise here; (c) Spectrum image of Gaussian fitted Mn-$L_3$ peak position of the interface area in (b).

The valence state of Mn was checked by doing nonlinear least square fitting of a Gaussian peak to the Mn $L_3$ peaks, as shown in Fig. 6. Spectra from the matrix region and the pillar center [Fig. 6(a)] show Mn $L_3$ peak positions of 639.75 eV and 638.25 eV, respectively, *i.e.* a difference of 1.5 eV. Comparison with literature data[27, 28] shows that this can be interpreted as a reduced valence state of Mn in the pillar. Applying the same procedure to the Mn spectrum image in Fig. 5(d) from the area in Fig. 6(b), which is a replica of Fig. 5(a), we obtain the Mn $L_3$ peak position map in Fig. 6(c), showing a decrease of Mn valence state from the matrix to the pillar.

Except the peak locations, the resulting Gaussian amplitude from non-linear least square fitting of La-$M_5$ [Fig. 7(b)] and Mn-$L_2$ [Fig. 7(c)] peaks reveals La and Mn positions within pillars at atomic resolution. As shown in Fig. 7, both of them occupy Zr positions within the pillar, which is highlighted by color-coded overlay of Fig. 7(b) and Fig. 7(c).

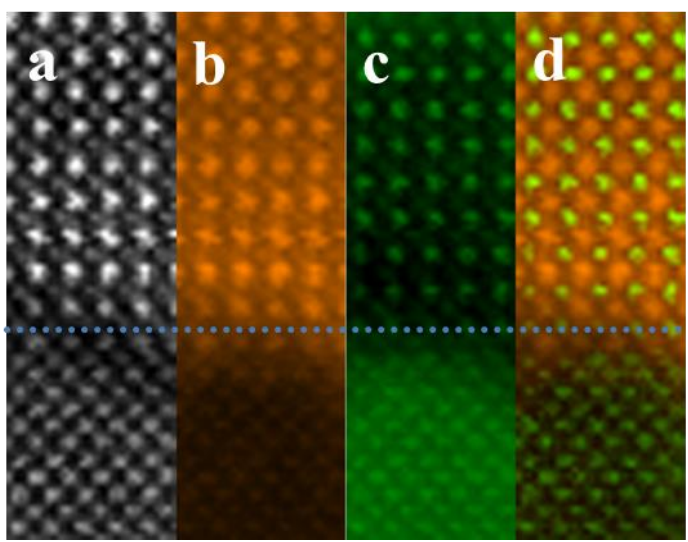

FIG. 7. (a) Same as Figure 5(a). (b) & (c) Spectrum image of Gaussian-fitted La-$M_5$ and Mn-$L_2$ peak amplitude, respectively. (d) Color overlay of (b) and (c).

The concentration profiles shown in Fig. 5 show enhanced La and Mn concentrations close to the left interface of the left pillar. We note here that this is not well pronounced in between close pillars. We discuss here only the left-hand interface because this resembles best the interface of an isolated pillar in the LSMO matrix. Sr is slightly depleted in this interface region whereas Zr is almost absent there. It seems that La occupies Sr positions close to the interface and the overall interface composition resembles $LaMnO_{3-x}$. As shown in Fig. 6, the Mn valence at the interface is decreased as compared to the LSMO matrix, probably to a value between 2 and 2.5. Because the La valence is 3+, charge balance can be obtained by introducing oxygen vacancies into the interface region. A loss of oxygen in this region is indeed shown in Fig. 5(l). Assuming a Mn valence of 2.5+, $x$ would be 0.25, corresponding to an oxygen vacancy concentration of 8 %.

It is known that Mn can stabilize the tetragonal or cubic structure of $ZrO_2$ at low temperatures by varying Mn concentrations.[29-31] Our observation of tetragonal or cubic $ZrO_2$ pillars and the presence of Mn within the $ZrO_2$ pillars indicates that this stabilization is a driving force for Mn dissolution within the pillars. This has also been observed by other groups in the past, e.g. [30, 32] For charge balance reasons oxygen vacancies have to accompany the $Mn_{Zr}$ substitution. This fits well with our observation of a loss of oxygen in the interface region where the concentration of $Mn_{Zr}$ is particularly high. The incorporation of oxygen vacancies permits stabilized zirconia to conduct $O^{2-}$ ions, provided there is sufficient vacancy mobility, a property that increases with temperature.

As a consequence $Mn_{Zr}$ substitution, the replaced Zr ions diffuse into the LSMO matrix. The detection of Zr in LSMO is consistent with former results,[33] where Zr-substituted $La_{0.7}Sr_{0.3}Mn_{1-x}Zr_xO_3$ with $0 < x < 0.20$ was investigated by neutron diffraction revealing that substitutional $Zr^{4+}$ occupies Mn site. $Zr^{4+}$ most likely replaces $Mn^{4+}$ ions [33], because of charge balance. However, one has to notice that the ionic radius of $Zr^{4+}$ is 0.72 Å whereas that of $Mn^{4+}$ is only 0.53 Å. This introduces strain and increases the lattice parameter of LSMO.[14] This imposes an upper limit to the soluble Zr amount. At concentrations above this solubility limit $ZrO_2$ precipitates nucleate, leading to a reduction of strain because of the smaller lattice parameter of the $ZrO_2$ precipitates compared to the LSMO matrix. Strain is then mainly localized at the precipitate−matrix interface.

Fig. S3 shows that $ZrO_2$ precipitates form even in the specimen with 6 mol% $ZrO_2$. This shows that the solubility of Zr in LSMO must be less than 6 mol%. There are literature data claiming higher solubility, e.g. [33, 34], but this is probably due to the higher processing temperature in these studies. As noted above the limited solubility may partially be due to the strain imposed by the large Zr ions in the LSMO matrix. Moreover, as mentioned by Kim *et al.*,[33] oxygen deficiency and interdiffusion may also play an important role. Oxygen vacancies have been found and studied before in LSMO.[35, 36] The concentration of oxygen vacancies are correlated with the sample preparation temperature and oxygen pressure. The existence of oxygen vacancies would convert some $Mn^{4+}$ to $Mn^{3+}$, thus reduce the number of $Mn^{4+}$ positions that can be replaced with $Zr^{4+}$. This is clearly confirmed by our results shown in Fig. 4 and 5: Zr reaches a minimum in the LSMO accompanied by oxygen deficiency. Therefore, the combination of charge balance, strain, and oxygen vacancies is the possible cause for the low Zr solubility in the matrix. These results are relevant for the equilibrium constitution of the $ZrO_2$/Zr/LSMO system.

Discussion of the implications of the observed microstructure for electron transport properties is beyond the scope of the present article. Probably these implications are manifold. Most importantly, the pillars act as scattering centers modifying the phase of the scattered electron wave function which can give rise to effects such as weak localization. Here, pillar size and densities play an important role. However, there might also be more subtle influences by the atomic substitutions mentioned above. For example, it is known that in LSMO electron transport is influenced by the double-exchange mechanism between Mn atoms of different valence ($Mn^{3+}$-O-$Mn^{4+}$). The substitution of Mn with Zr can therefore be expected to influence this mechanism and thus the electronic transport in this system.

## IV CONCLUSIONS

In summary, we have presented atomic-scale studies of the structure and chemistry of the $ZrO_2$−LSMO pillar matrix system. We showed that $ZrO_2$ precipitates form at as low concentrations as 6 mol%. Precipitates mainly form pillars penetrating the entire LSMO film. Substantial interdiffusion is found at the LSMO−$ZrO_2$ interface with Mn replacing Zr in $ZrO_2$ (thus stabilizing the cubic or tetragonal phase) and Zr replacing Mn atoms in LSMO. Charge balancing requires formation of oxygen vacancies which are observed to segregate at the interface. Strain analyses show that the system has not yet reached elastic equilibrium. It is clear that LSMO as well as pillar regions are strained because of the misfit, however, also modified by the interdiffusion. In the $La_{(1-y)}Sr_yMnO_3$ system, the magnetic properties are directly related to the $y$ value.[37] Therefore, we believe that our results are relevant for the underlying mechanisms of electron transport and magnetism in this material system, like the observed anomalous transport properties and localization transition.

## SUPPORTING INFORMATION

Supporting Information, including structure model and more TEM results, Figure S1-S3, is available online or from the author.

## ACKOWLEDGEMENT

Planck Institue for Solid State Physcis (MPI-FKF). The research leading to these results has received funding from the European Union Seventh Framework Program [FP/2007/2013] under grant agreement no 312483 (ESTEEM2).